\documentclass[10pt]{iopart}

\usepackage{graphicx}
\usepackage{epstopdf} % Remove before submission, at least the outdir option
\graphicspath{{}} % Set a custom graphics path, remove before submission
\usepackage[utf8]{inputenc}
\usepackage[T1]{fontenc}
\usepackage{subfig}
\usepackage[numbers,sort&compress]{natbib}
\usepackage{hyperref} 
	\hypersetup{colorlinks=true,
		pdfstartview=FitV,
		linkcolor=blue,
		citecolor=blue,
		urlcolor=blue,
	}
\usepackage{hypernat}

\let\oldsim\sim 
\renewcommand{\sim}{{\oldsim}}

\begin{document}

\title[Runaway dynamics in argon-induced disruptions]{Modelling of runaway electron dynamics during argon-induced disruptions in ASDEX Upgrade and JET}

\author{K.~Insulander Bj\"{o}rk$^1$,
  O.~Vallhagen$^1$,
  G.~Papp$^2$,
  C.~Reux$^3$,
  O.~Embreus$^1$,
  E.~Rachlew$^1$,
  T.~F\"{u}l\"{o}p$^1$,
  the ASDEX Upgrade Team\footnote{See the author list of ``H. Meyer {\em et al}, 2019 Nucl. Fusion \href{https://iopscience.iop.org/article/10.1088/1741-4326/ab18b8}{{\bf 59} 112014}''},
  JET contributors\footnote{See the author list of ``E. Joffrin {\em et al}, 2019 Nucl. Fusion \href{https://iopscience.iop.org/article/10.1088/1741-4326/ab2276}{{\bf 59} 112021}''} and
  the EUROfusion MST1 Team\footnote{See the author list of ``B. Labit {\em et al}, 2019 Nucl. Fusion \href{https://iopscience.iop.org/article/10.1088/1741-4326/ab2211}{{\bf 59} 086020}''}}

\address{$^1$Chalmers University of Technology, Gothenburg, 412 96, Sweden}
\address{$^2$Max Planck Institute for Plasma Physics, D-85748 Garching, Germany}
\address{$^3$CEA, IRFM, F-13108 Saint-Paul-lez-Durance, France}
\ead{klaraib@chalmers.se}

\begin{abstract}
  Disruptions in tokamak plasmas may lead to the generation of runaway electrons that have the potential to damage plasma-facing components. Improved understanding of the runaway generation process requires interpretative modelling of experiments. In this work we simulate eight discharges in the ASDEX Upgrade and JET tokamaks, where argon gas was injected to trigger the disruption. We use a fluid modelling framework with the capability to model the generation of runaway electrons through the hot-tail, Dreicer and avalanche mechanisms, as well as runaway electron losses. Using experimentally based initial values of plasma current and electron temperature and density, we can reproduce the plasma current evolution using realistic assumptions about temperature evolution and assimilation of the injected argon in the plasma. The assumptions and results are similar for the modelled discharges in ASDEX Upgrade and JET. For the modelled discharges in ASDEX Upgrade, where the initial temperature was comparatively high, we had to assume that a large fraction of the hot-tail runaway electrons were lost in order to reproduce the measured current evolution.

\end{abstract}


\submitto{Plasma Physics and Controlled Fusion}
\maketitle
\ioptwocol

\section{Introduction}
\label{sec:intro}

\textit{Runaway electrons} (RE) may cause severe damage to plasma-facing components in tokamaks \cite{Hender_2007}, where they may occur as a consequence of disruptions. To avoid this, different schemes are being developed to mitigate, limit or entirely avoid the formation of REs. Massive material injections (MMI) are proposed to this end, which may be realized through a gas injection (massive gas injection - MGI) or the injection of solid pellets, which can be shattered when entering the vacuum chamber (shattered pellet injection - SPI) \cite{HollmannDMS, lehnen15disruptions}. In medium-sized tokamaks, the potential of these measures have been demonstrated \cite{hollmann15measurement, Pautasso_2016, pazsoldan17spatiotemporal, carnevale18runaway, reux15runaway, esposito17runaway, coda2019physics, mlynar19runaway, pautasso20generation,Linder-2020,linder2021electron}, but in the future devices envisaged for fusion energy generation, the plasma conditions will be significantly different (higher temperatures and densities, larger plasma currents), and whether the proposed measures for RE mitigation or avoidance will be effective also in these devices can currently only be assessed through plasma physics modelling.

Several theoretical models for the physics of runaway electrons in tokamak disruptions  have been developed and implemented in computational tools \cite{BreizmanAleynikov2017Review}. To assess to which extent these models can be applied to make useful predictions for the plasma behaviour during disruptions, they must be validated against existing experimental data. In current tokamaks, disruptions are often deliberately triggered by MGI, leading to a thermal quench (TQ) which under certain conditions is followed by a formation of REs and a rapid decay of the ohmic plasma current (a current quench - CQ). The experimental data collected during such discharges constitutes a valuable dataset for validation of models for the formation of REs in the presence of impurities (often in the form of massive amounts of noble gasses). In this work, we use such experimental data from the tokamaks  ASDEX Upgrade (AUG) and Joint European Torus (JET)  to investigate the applicability of the models, possible modifications needed, and qualitative differences between the mechanisms behind RE formation in these two differently sized tokamaks.

The main computational tool used in the present paper is called \textsc{go}, a fluid code that describes the radial dynamics of the current density and the electric field, in the presence of impurities. The models implemented in this tool are briefly described in section \ref{sec:models}, and more details are given in Refs.~\cite{Smith-PoP-06,Feher-PPCF-11,Papp-NF-13,Vallhagen-JPP-20}. In the version of {\sc go} that is used in the paper, hot-tail, Dreicer and avalanche RE generation models are implemented, as will be further described in section \ref{sec:models}.

REs are defined as electrons having a momentum larger than the \textit{critical momentum} $p_c$ (or, equivalently, a velocity larger than the critical velocity $v_c$), above which the accelerating force exerted by the electric field (induced in the disruption) is larger than the friction force due to collisions, \textit{i.e.}~electrons above this threshold are accelerated until their momentum is limited by radiative energy losses. When the collision time at the critical momentum is longer than the duration of the TQ, hot-tail is the dominant primary generation mechanism \cite{helanderhottail}. In future fusion devices, this is expected to be the case \cite{SmithHottail2005}, so significant hot-tail generation is expected  in e.g.~ITER \cite{martinsolis17formation} or SPARC \cite{Sweeney2020}. For this reason, it is important to properly understand and being able to model the interplay of the various runaway generation mechanisms in experimental scenarios.

Kinetic simulations of AUG discharges, using the tool \textsc{code} \cite{Stahl-NF-2016}, were recently presented by Insulander Bj\"{o}rk {\em et al.}~\cite{Insulander-JPP-20}. \textsc{code} solves the linearized Fokker-Planck equation including radiation reactions and avalanche source, as well as the electric field evolution \cite{Ecrit}. Kinetic simulations of JET discharges with {\sc code} were attempted, but became prohibitively slow due to the high electric fields induced. Also, \textsc{code} lacks radial diffusion of electric field and current, mechanisms which have been shown to be important for modelling of disruptions in large machines such as ITER \cite{Vallhagen-JPP-20}, so in this sense, the two codes are complementary.

%In this paper, we begin by comparing simulations of AUG discharges with \textsc{go} and \textsc{code} and verify that these tools make qualitatively similar predictions, despite the different models used.
In this paper, we use the \textsc{go} code to investigate argon-induced disruptions in the AUG and JET tokamaks.
In agreement with recent results based on coupled fluid and kinetic
simulations with \textsc{go} and \textsc{code}
\citep{Hoppe-ASDEX-2020}, we find that taking into account all the
hot-tail electrons in AUG would overestimate the final runaway current. To
be able to match the experimentally obtained current evolution, we
vary the loss-fraction of the hot-tail RE. Such losses are expected to
occur due to the breakup of magnetic flux surfaces, accompanying the TQ. 

Furthermore, we find that the self-consistent calculation of the
temperatures of ions and electrons using time-dependent energy
transport equations often results in a temperature evolution which
does not agree with measurement data. The main reason why the energy transport equations fall short of accurately predicting the temperature evolution is likely the lack of a detailed self-consistent model for the three-dimensional evolution of the magnetic field during the TQ. However, by assuming an
exponentially decaying temperature evolution throughout the
simulation, the experimentally observed current evolution can be
reproduced both in AUG and JET.

%%%%%%%%%%% Modelling
\section{Fluid modelling with \textsc{go}}
\label{sec:models}

{\sc go} \cite{Smith-PoP-06,Feher-PPCF-11,Papp-NF-13,Vallhagen-JPP-20} simulates the radial dynamics of the of the current, temperature, density and electric field. Models for the previously mentioned RE generation mechanisms are implemented in the code, as well as RE generation through Compton scattering and tritium decay, albeit the two latter mechanisms are not relevant for the simulations presented here.

\subsection{Runaway electron generation}
The \textit{hot-tail} RE generation is a result of a rapid cooling of the plasma, during which there is not enough time for the fastest electrons (the "hot tail" of the electron velocity distribution) to thermalize, thereby ending up above the critical momentum. The analytical model used in {\sc go} for hot-tail generation is derived in Ref.~\cite{Smith-PoP-08}.
\begin{eqnarray}
       \frac{d n_{\rm RE}^{h}}{dt} = -\frac{du_c}{dt} \frac{2 u_c^2H(-du_c/dt)}{(u_c^3-3 \tau)^{1/3}}\int_{u_c}^\infty\frac{e^{-u^2}u^2 du}{(u^3-3 \tau)^{2/3}},
    \label{eq:hottail}
\end{eqnarray}
where $u_c=(v_c^3/v_{T0}^3+3 \tau)^{1/3}$, $\tau=\nu_0\int_0^t
n(t)/n_0 dt$, $n$ is the electron density, $v_{T}$ is the thermal
electron speed, $v_c$ is the critical velocity, subscript $0$ denotes an
initial value, $H$ is the Heaviside function, $\nu=n e^4
\ln\Lambda/(4\pi \epsilon_0^2 m_e^2 v_T^3)$, ln$\rm \Lambda$ is the Coulomb logarithm, $\epsilon_0$ is the vacuum permittivity and $m_e$ the electron rest mass. This expression takes
into account the directivity of the electric field, in contrast to the simplified
expression used in Refs.~\cite{Fulop2009,martinsolis17formation}, which is based on counting the number of electrons with $v>v_c$.

The \textit{Dreicer} RE generation is instead a consequence of momentum space diffusion of electrons to momenta above the critical threshold, due to small angle collisions \cite{dreicer1959}. For modelling of Dreicer RE generation, a neural network was used, trained on the output of kinetic simulations of plasmas with impurities, which were performed with \textsc{code} \cite{Hesslow-JPP-19}.

RE generation through these two mechanisms is referred to as \textit{primary generation}. Already existing REs thus generated may transfer part of their momentum to bulk electrons, knocking them above the critical momentum and thereby creating new REs at an exponentially growing rate. This a \textit{secondary generation} mechanism, referred to as \textit{avalanche}  \cite{sokolov1979multiplication,RosenbluthPutvinski1997}. The avalanche generation is modelled by a semi-analytical formula \cite{Hesslow-NF-19}, that has been carefully benchmarked against kinetic {\sc code} simulations of disruptions in impurity-containing plasmas.

\subsection{Temperatures}
The temperature evolution in this work was simulated using different models, described below. 

\subsubsection{Exponential temperature model}
\label{sec:GOtemperatureExp}
During the initial phase of the simulations when the temperature decays rapidly (TQ), the magnetic flux surfaces are broken up due to magneto-hydrodynamic (MHD) instabilities \cite{BoozerMSloss}. The rapid radial transport due to these effects is the main driving force behind the initial temperature change. This complex transport process cannot be self-consistently modeled within our fluid framework, and emulating it in terms of spatiotemporally varying transport coefficients would introduce a large set of unconstrained free parameters. Instead, in our simulations the on-axis temperature $T_{\rm oa}$ is initially prescribed to follow an exponential decay \cite{Smith-PoP-08, martinsolis17formation}:
\begin{equation}
T_{\rm oa} = T_{\rm CQ} + (T_{\rm initial}-T_{\rm CQ}) \cdot e^{-\frac{t}{t_{\rm TQ}}},
\label{eq:Texp}
\end{equation}
where  the initial temperature $T_{\rm initial}$ is deduced from experimental data, and the thermal quench time $t_{\rm TQ}$ and the temperature at the end of this phase and most of the CQ, $T_{\rm CQ}$, may be treated as free parameters which can be adjusted to yield an $I_{\rm p}$ evolution similar to the measured data. Another reason for using this prescribed temperature evolution during the TQ is to be consistent with the assumptions used to derive the analytical model for the hot-tail runaway generation \cite{Smith-PoP-08}. During this initial temperature drop, the radial temperature profile shape is assumed to remain constant in the present simulations. The initial radial temperature profile $T_{\rm initial}$ is deduced from experimental data as described in sections \ref{sec:AUGtemperature} and \ref{sec:JETtemperature}, for AUG and JET respectively, and this distribution is then scaled with the value given by equation \eref{eq:Texp} to calculate the free electron temperature in each radial point. Whereas this is undoubtedly a coarse approximation, experimental measurements cannot yield useful data on the actual temperature profile during this brief and turbulent phase of the disruption to support a better estimate. Assuming a flat temperature profile with a value  corresponding to a homogeneous spreading of the initial thermal energy will lead to slightly higher final runaway currents, but do not lead to qualitative changes in the results. Note, that since $T_{\rm CQ}$ is assumed to be constant, the final temperature profile is flat in all simulations. 

\subsubsection{Energy transport temperature model}
\label{sec:GOtemperatureEq}
\textsc{go} also has the capability to calculate the temperatures of free electrons, bulk ions and impurity ions using time-dependent equations for the energy transport of the respective species in a one-dimensional cylindrical model \cite{Feher-PPCF-11}. Whereas these equations do not account for particle transport, and are not sufficiently well constrained during the TQ, an attempt  was made to use them for calculation of the post-TQ temperatures. The energy transport equations are implemented as
\begin{eqnarray}
    \frac{3}{2}\frac{\partial  (n T)}{\partial t}&=\displaystyle\frac{3
      n}{2 r}\frac{\partial
    }{\partial r}\left(\chi r \frac{\partial T}{\partial
        r}\right)+\sigma(T, Z_{\rm eff})E^2\nonumber \\&-\sum_{i,k}n_\mathrm{e}n_{k}^iL_{k}^i(T,n_e)-P_{\rm Br}-P_{\rm ion},     
    \label{eq:ebalance}
\end{eqnarray}
where $n$ is the total density of all species (electrons and ions), $\sigma$ is the conductivity, $Z_{\rm eff}$ is the effective ion
charge, $n^k_{i}$ is the density of the $i^{\rm th}$ charge state ($i=0, 1, ..., Z-1$) of the ion species $k$ (e.g.~deuterium, argon), $P_{\rm Br}$ and $P_{\rm ion}$ are the Bremsstrahlung and the ionization energy losses, respectively. The line radiation rates $L_k^{i}(T,n_e)$ are extracted from ADAS \cite{ADAS}. In these calculations, which were only attempted for the post-TQ plasma, a constant heat diffusion coefficient $\chi=1\,\rm m^2/s$ was used, and it was confirmed that the results were insensitive to this choice. %When using the energy transport temperature model, the temperature profile changes in accordance with the model.

In these simulations, the temperature is assumed to be equal for all species, since this is computationally less expensive. This simplification affects the heat capacity with a factor of at most two \cite{Vallhagen-JPP-20}. Since the initial part of the temperature evolution, where most of the thermal energy is lost, is not simulated with this model, this simplification is not expected to have an important impact. The injected argon is instead assumed to be heated through  heat exchange with the particles present before the injection.

\subsubsection{Hybrid temperature model}
\label{sec:GOtemperatureHyb}
An attempt was made to use a hybrid temperature model. First, the temperature was calculated using equation \eref{eq:Texp} with $T_{\rm CQ}$ = 1 eV, until $T_{\rm oa}$ reached a pre-set value $T_{\rm switch}$. It has been estimated \cite{Ward_Wesson} that transport due to stochastic flux surfaces could drive a temperature drop down to $\sim 100$ eV more effectively than radiation, which provides a physically motivated value for $T_{\rm switch}$. Then the simulation was restarted using the time-dependent energy transport equation \eref{eq:ebalance} for temperature determination, i.e.~the value $T_{\rm CQ}$ = 1 eV is not reached. After the switch to using the energy transport equations the temperature is calculated locally in each radial point by \textsc{go}. 

In many cases, this strategy resulted in a rapid increase in the calculated on-axis temperature up to a few 100 eV after the switch. Re-heating has been sporadically observed in natural disruptions \cite{Pautasso03Study}, but has not been possible to investigate experimentally in MGI-induced disruptions. On ASDEX Upgrade, the electron cyclotron emission temperature diagnostic (see section~\ref{sec:AUGtemperature}) is in density cut-off after the MGI, therefore any potential re-heating can not be directly measured. However, it is likely that the losses in the energy transport equation are underestimated, for example due to radiation from wall impurities or remaining transport losses that are not included, or underestimated, in the present model. The radiation losses with argon impurities present follow a non-monotonic behaviour as a function of temperature, with maxima at $\sim 10$ eV and $\sim 100$ eV. Even a small underestimate of the radiative losses can make the ohmic heating overcome the cooling in the $\sim 10$ eV range, and make the temperature evolve towards the $\sim 100$ eV range where the ohmic heating is less efficient. However, if the calculated losses are large enough, the temperature will instead move towards the $\sim 10$ eV range, where it remains until the ohmic heating has decayed. When the simulations result in such a temperature evolution also the experimental $I_{\rm p}$ evolution was reproduced fairly well.

There is a close connection between the current decay rate and the temperature through the conductivity. In cases where re-heating occurred in the simulations, the experimentally observed current decay rate was not reproduced sufficiently well. Thus, in sections \ref{sec:Ipmatching} and \ref{sec:JETexponential} we used the exponential temperature model only, i.e.~equation \eref{eq:Texp} was used for the entire simulation, and the temperature $T_{\rm CQ}$ was adjusted, together with other free parameters, to match the measured $I_{\rm p}$ evolution. The values of $T_{\rm CQ}$ which could match the $I_{\rm p}$ evolution were approximately 20 eV, i.e.~$T_{\rm CQ}$ is the free electron temperature during the most of the CQ, but not necessarily the final temperature during the RE plateau phase.

\subsection{Densities}
\label{sec:GOdensities}
\textsc{go} calculates the ionization states of the argon and the resulting free electron density from the time-dependent rate equations
\newcommand{\ud}{\mathrm{d}}
\begin{equation}
    \frac{\ud n_{k}^i}{\ud t}=  n_{\rm e} \left[I_k^{i-1} n_{k}^{i-1} - (I_k^i+ R_k^i) n_{k}^{i} +  R_k^{i+1} n_{k}^{i+1} \right],
 \label{eq:tdre}\end{equation}
using as input the free electron density prior to the injection, the injected amount of argon and the initial density profile. $I_k^{i}$ denotes the electron impact ionization rate and $R_k^i$ the radiative recombination rate for the $i^{\rm th}$ charge state of species $k$, respectively. The ionization and recombination rates are extracted from ADAS \cite{ADAS}. The injected argon is assumed to distribute according to the same density profile as the initial deuterium density.

In the current simulations, the quantity of assimilated argon is given in terms of the ratio $r_{\rm Ar/D}$ of the argon density $n_{\rm Ar}$ to the initial deuterium density, assumed to be equal to the initial free electron density $n_{\rm e0}$.
Lacking reliable experimental data on the rate of assimilation of argon into the plasma, the argon is assumed to have assimilated and distributed within the plasma before the beginning of the simulated time span. The density of free electrons and of each ionization state is calculated by \textsc{go}. $T_{\rm oa}$, calculated by equation \ref{eq:Texp}, is the on-axis free electron temperature including cooling by the injected argon. The effect of modelling the argon density as exponentially increasing after the beginning of the simulation was investigated, but this introduced another free parameter (the time rate of argon assimilation) which affected the simulation results in a similar manner as assuming a constant $r_{\rm Ar/D}$, so the simpler approximation was preferred.

The evolution of the free electron density during the disruption is difficult to diagnose directly. For AUG, the total assimilated amount of argon was deduced from the current dissipation rate during the RE plateau phase, as explained in Ref.~\cite{Insulander-JPP-20}, and the same assimilation rates were used in the present simulations. For JET, experimental data was used to determine the total amount of assimilated argon, as explained in section \ref{sec:JETexponential}. However, we did not attempt to deduce the time evolution of the argon assimilation from these data, partly due to their uncertain nature, and partly in order to use the same modelling strategy for both tokamaks.

\subsection{Plasma current}
\label{sec:GOcurrent}
{\sc go} models the evolution of the current density as a balance between the diffusion of the electric field and the generation of the REs:
\begin{equation} \frac{1}{r}\frac{\partial }{\partial r}
  r\frac{\partial E_\|}{\partial r}=\mu_0 \frac{\partial j_\parallel}{\partial
    t},
\label{GO}
\end{equation} where 
 $j_\parallel= \sigma_{\rm Sp} E_\|+ecn_{\rm RE}$ is the sum of Ohmic
and runaway current densities, with $\sigma_{\rm Sp}$ being Spitzer
conductivity, and $\mu_0$, $e$, and $c$ denote the magnetic permeability,
the elementary charge and the speed of light,
respectively. Toroidicity effects are neglected here.

In the disruptions modelled in this paper, the measured plasma current $I_{\rm p}$ typically displays a small peak coinciding with the beginning of the TQ, which indicates an MHD event during which the current density is redistributed radially due to a breakup of magnetic flux surfaces \cite{BoozerMSloss}. Towards the end of the TQ, the magnetic flux surfaces re-form and the current can be confined until it is dissipated resistively.

The rapid relaxation of the current profile due to non-axisymmetric MHD is not modelled by \textsc{go}, and thus the calculated $I_{\rm p}$ does not display this peak. After the peak, $I_{\rm p}$ starts to decay rapidly (the CQ) due to the increase in plasma resistivity associated with the decreasing temperature ($\sigma_{\parallel} \propto T_e^{3/2}$), which is modelled by \textsc{go}. Due to this temperature dependence, the $I_{\rm p}$ evolution is strongly affected by $T_{\rm e}$ evolution, and hence by the parameters describing it: $T_{\rm initial}$, $T_{\rm CQ}$ and $t_{\rm TQ}$. As previously stated, the exact value of $T_{\rm CQ}$ does not affect the results significantly if we use the hybrid temperature model, i.e.~if we switch to the time-dependent energy transport equation before $T_{\rm CQ}$ is reached. However, the value becomes important when we use the exponential decay temperature model, i.e.~if equation \eref{eq:Texp} is used throughout the simulation, as discussed in sections \ref{sec:AUGJETcomparison} and \ref{sec:Ipmatching}. In these cases, $T_{\rm CQ}$ rather plays the role of the equilibrium temperature that marks the end of the TQ, but not necessarily the final temperature reached after the CQ when no ohmic current remains causing resistive heating of the plasma.

\subsection{Plasma elongation}
\label{sec:GOelongation}
\textsc{go} has the ability to model elongated plasmas \cite{Fulop2020}, however, this capacity was not used in the presented simulations. The modelled discharges featured slightly elongated plasmas, $\kappa \approx$ 1.16 in the JET discharges, according to pre-disruption equilibrium reconstructions\footnote{EFIT/ELON}. Inclusion of the elongation in the simulations did not result in any qualitative changes in the simulation results, and thus it was decided to reduce the parameter space by omitting this parameter. This simple geometry also facilitates comparison with previous kinetic simulations which were zero-dimensional in real space.

%%%%%%%%%%% Experimental data %%%%%%%%%%%%%%%%%%%%%%%%%%
\section{Experimental data}
\label{sec:experimental}

In order to assess the applicability of the \textsc{go} model for tokamaks with different parameter ranges, we model discharges in AUG and JET, and compare our results to experimental data. Table \ref{tab:tokamaks} shows some basic data typical for the discharges simulated for the respective tokamaks, and in the following sections, more specific data for each simulated discharge are presented.

\begin{table*}
\centering
\caption{Typical parameters for the modelled discharges in the respective tokamak. $^*$At the time of the modelled disruptions, the minor radius was 0.8 m.}
\label{tab:tokamaks}
\begin{tabular}{l c c} 
\br
Tokamak 												& AUG		& JET		\\
\mr
Major radius	$R$ [m]										& 1.65		& 3.0		\\ 
Minor radius	$a$ [m]										& 0.50		& 1$^*$		\\ 
Initial plasma current $I_{\rm p0}$ [MA]					& 0.76		& 1.3 - 1.4	\\
Initial on-axis free electron temperature $T_{\rm initial}$ [keV]		& 5.3 - 7.2 	& 1.6 - 2.0 	\\
Initial on-axis free electron density $n_{e0}$ [$10^{19}$ m$^{-3}$] 	& 2.4 - 3.1 	& 4.5 - 8.7 	\\
Toroidal magnetic field on axis $B$ [T] 						& 2.5 		& 3 			\\
\br
\end{tabular}
\end{table*}

% ASDEX Upgrade %%%%%%%%%%%%%%%%%%%%%%%%%%
\subsection{ASDEX Upgrade} \label{sec:AUG}
ASDEX Upgrade is a medium sized tokamak, and we model four discharges which were specifically designed for the study of runaway electron dynamics \cite{Pautasso_2016,pautasso20generation}. The runaway discharges studied in this paper are near-circular (elongation $\kappa \approx 1.1$), inner wall limited, have core ECRH (Electron Cyclotron Resonance Heating)  and low pre-disruption density ($n_{\rm e0}\approx 3{\cdot}10^{19}~\mathrm{m}^{-3}$). The discharges are terminated using argon MGI from an in-vessel valve. A 30 ms time period, starting at the argon valve trigger, was simulated to ensure that the entire current quench was covered in all the simulated cases. The plasma position remained radially and vertically stable in the experiments during the simulated time window. 

Four discharges with different plasma parameters and amounts of injected argon were selected for modelling and an overview of the basic parameters for these discharges is presented in table \ref{tab:AUGshots}. These discharges were selected from the set of 11 discharges modelled earlier in Ref.~\cite{Insulander-JPP-20}.

\begin{table*}
\centering
\caption{Basic parameters for the four simulated discharges in AUG, and the notation for these parameters used in this paper. Further descriptions of the origins of these parameters are found in sections \ref{sec:AUGtemperature} and \ref{sec:AUGdensities} below.}
\label{tab:AUGshots}
\begin{tabular}{l c c c c} 
\br
Discharge number												& 33108	& 34183	& 35649	& 35650 \\
\mr
Injected number of Ar atoms $N_{\rm{Ar}}$ [$10^{19}$]			& 175	& 74	& 94	& 96	\\
Initial free electron temperature on axis  $T_{\rm initial}$ [keV]				& 7.2	& 5.5	& 6.2	& 5.3	\\			
Initial on-axis free electron density $n_{\rm e0}$ [$10^{19}$ m$^{-3}$]	& 3.1	& 2.8	& 2.6	& 2.4	\\
\mr
\multicolumn{5}{l}{\textit{Parameters used in the initial \textsc{go} simulations}}\\
\mr
Thermal quench time parameter $t_{\rm TQ}$ [ms]					& 0.152	& 0.198	& 0.178	& 0.177	\\
Argon-to-deuterium density ratio $r_{\rm Ar/D}$					& 1.40	& 0.64	& 0.90	& 0.98	\\
\br
\end{tabular}
\end{table*}

\subsubsection{Temperature} \label{sec:AUGtemperature}
The temperature and its radial distribution in the plasma is measured using electron cyclotron emission (ECE). For the present simulations, we only use the initial temperature profile. The initial on-axis temperature $T_{\rm initial}$ used in equation \eref{eq:Texp} is determined as the average temperature over the circular area within 1 cm of the magnetic axis and the time interval between 1 ms and 1.5 ms after the argon injection valve trigger. The time interval was selected to exclude both the beginning of the TQ and any initial temperature variations due to the shut-off of the heating system shortly before the disruption. 

\subsubsection{Densities} \label{sec:AUGdensities}
The free electron density is measured by CO$_2$ interferometry, which yields the line integrated free electron density along the line of sight of the interferometer. The density profile is fitted by the tool \textsc{augped}, which fits a modified hyperbolic tangent function \cite{Schneider-thesis-12} to the radial density data points obtained with interferometry. The average measured free electron density during the first 1.5 ms after the argon valve trigger is used as the initial on-axis free electron density $n_{\rm e0}$, since after these 1.5 ms, the argon injection causes the measured density to start to increase. The initial density profile is scaled with $n_{\rm e0}$ to obtain the initial density in each radial data point. 

In each discharge, argon gas was injected into the vacuum vessel to trigger the disruption \cite{Pautasso_2016, Fable-NF-16}. The injected number of argon atoms $N_{\rm{Ar}}$ is calculated from the pressure in the MGI reservoir holding the argon gas before injection, the reservoir volume (0.1 l) and the gas temperature (300 K). This quantity is listed in table \ref{tab:AUGshots} for the respective discharges. The amount of argon which finally assimilates in the plasma as a function of time is difficult to assess. In previous work \cite{Insulander-JPP-20}, the Ar assimilation fraction $f_{\rm Ar}$ was assumed to be the same for all discharges, and was assessed by matching the current decay rate during the RE plateau phase. It was found that a reasonable estimate was that by the end of the TQ, 20\% of the injected argon was assimilated in the plasma volume $V_{\rm p}$, as defined by the minor and major radii listed in table \ref{tab:tokamaks}. We hence make the same assumption for the present simulations, and calculate the argon density $n_{\rm Ar}$ as $N_{\rm{Ar}}/V_{\rm p}\cdot f_{\rm Ar} = N_{\rm{Ar}}/V_{\rm p}\cdot 0.2$. The ratio $r_{\rm Ar/D}$ is calculated as $n_{\rm Ar}/n_{\rm e0}$. 

% JET %%%%%%%%%%%%%%%%%%%%%%%%%%
\subsection{Joint European Torus - JET}
\label{sec:JET}
JET is significantly larger than ASDEX Upgrade, as shown in table \ref{tab:tokamaks}. Four discharges in which runaway electrons were formed were chosen for modelling, representing a set of varying plasma parameters. In particular, the size of the Ar injection varies by more than an order of magnitude. The basic parameters for these discharges are presented in table \ref{tab:JETshots}. 

\begin{table*}
\centering
\caption{Basic parameters for the four simulated discharges in JET and the notation for these parameters used in this paper. $^*$In discharges  \#92454 and \#92460, the Ar injection was succeeded by a large Kr injection. However, the effect on the $I_p$ evolution is insignificant until a few ms after the end of the modeled CQ.}
\label{tab:JETshots}
\begin{tabular}{l c c c c c} 
\br
Discharge number													& 92454$^*$	& 92460$^*$	& 95125	& 95129  \\
\mr
Injected number of Ar atoms $N_{\rm{Ar}}$ [$10^{19}$]        			& 882	& 147			& 75	& 74\\ 
Initial on-axis free electron temperature $T_{\rm initial}$ [keV] 			& 1.9	& 1.6			& 2.0	& 1.9\\
Initial free electron density  $n_{\rm e0}$ [$10^{19}$ m$^{-3}$]			& 8.7	& 7.4			& 4.5	& 5.2\\
Initial plasma current $I_{\rm p}$ [MA]                             				& 1.4	& 1.4			& 1.3	& 1.3\\
\br
\end{tabular}
\end{table*}

\subsubsection{Temperature} \label{sec:JETtemperature}
The free electron temperature is measured by high resolution Thomson scattering (HRTS) which yields both a temperature profile and an sufficiently reliable value of the on-axis temperature before the Ar injection ($T_{\rm initial}$ used in equation \eref{eq:Texp}). The initial temperature profile is smoothed using the \texttt{rloess} algorithm in \textsc{matlab} \cite{matlab} to remove signal noise.% Equation \eref{eq:Texp} was used for the entire simulation, and the TQ time parameter $t_{\rm TQ}$ and the final temperature $T_{\rm CQ}$ were varied along with the other free parameters (see section \ref{sec:Ipmatching}) in order to find a set of parameters which reproduced the measured $I_{\rm p}$ evolution.

\subsubsection{Densities} \label{sec:JETdensities}
The free electron density $n_e$ is measured by several different diagnostics. Interferometry yields the line integrated free electron density along the line of sight of the interferometer. In these simulations, we used the fast interferometer signal\footnote{KG4C/LDE3 signal, compensated for fringe jumps and scaled to give a pre-disruption density in agreement with the DF/G1C-LD<DCN:003 signal.}. The shape of the initial free electron density profile was retrieved from the HRTS data at the last available data point for this diagnostic before the Ar injection, which was at most 130 ms before the injection. The profile was smoothed to remove signal noise using \texttt{rloess} \cite{matlab} just as for the temperature profile, and then scaled so that the integral of the profile yielded the same initial line integrated density as the interferometry based density data.

The final number of injected argon atoms is calculated from the volume of the reservoir holding the argon before the injection (DMV3, with volume 0.35 L), the gas vessel pressure before and after the injection\footnote{DE/Y8C-GPT2B<FST signal.}, and the assumption that the gas is held at room temperature (300 K). The fraction of this argon which actually assimilates in the plasma is difficult to assess experimentally, and the gas also leaves the injection reservoir during a period of some tens of ms. Hence, the ratio $r_{\rm Ar/D}$ was regarded as a free parameter, constrained by the condition that the calculated maximum line integrated free electron density during the simulated time span should not deviate by more than 10\% from the maximum line integrated free electron density given by the interferometry.

%%%%%%%%%%% Results
\section{Results} \label{sec:results}

\subsection{AUG simulations with the hybrid temperature model}
\label{sec:initial}

In the first set of simulations, four AUG discharges were simulated using the same modelling parameters $t_{\rm TQ}$ and $r_{\rm Ar/D}$ as earlier \cite{Insulander-JPP-20} to verify that the modelled quantities show a qualitatively similar behaviour for \textsc{go} and the kinetic tool (\textsc{code}) used in Ref. \cite{Insulander-JPP-20}. For these simulations, the hybrid temperature model described in section \ref{sec:GOtemperatureHyb} was used, i.e.~equation \eref{eq:Texp} was used initially, and then we switched to the time-dependent energy transport equation after the on-axis $T_{\rm e}$ had dropped to below a pre-defined temperature $T_{\rm switch}$. This method most closely resembles the strategy used in Ref.~\cite{Insulander-JPP-20}. We set $T_{\rm switch}$ = 6 eV, which is similar to the temperature equilibrium value found in several of the simulations in Ref.~\cite{Insulander-JPP-20}. Furthermore, we set $T_{\rm CQ}$ = 1 eV, and note that the value of $T_{\rm CQ}$ has a minor impact on the exponential function, as long as it is lower than $T_{\rm switch}$, since the exponential function is abandoned before $T_{\rm CQ}$ is reached.

\begin{figure}
(a)\vspace{-0.2cm}\\ \subfloat{\includegraphics[width=0.43\textwidth]{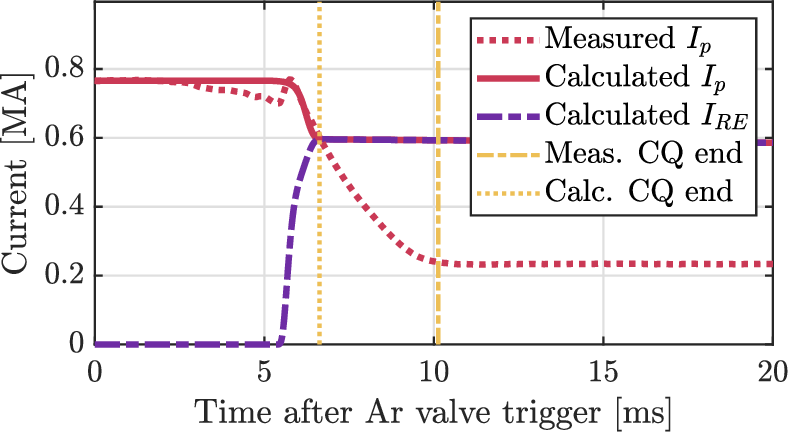}}
\vspace{-0.6cm}\\
(b)\\ \subfloat{\includegraphics[width=0.43\textwidth]{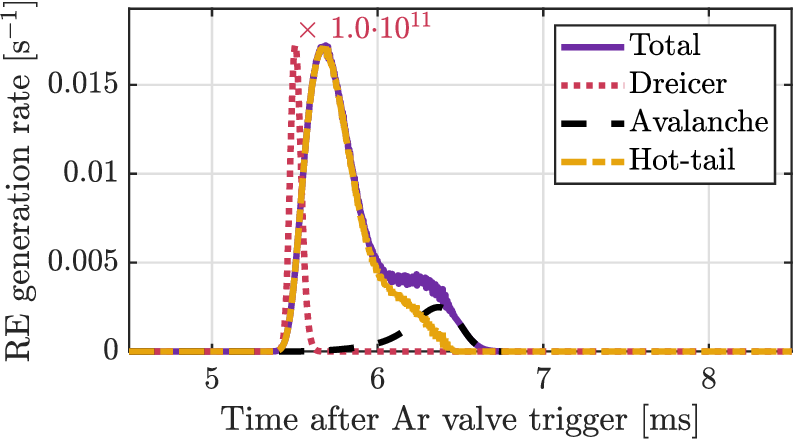}}
    \caption{(a) Current evolution and (b) RE generation rates for AUG discharge \#34183, using the hybrid temperature model. In panel (a), the experimentally measured total plasma current is plotted for reference. The vertical lines mark the end of the CQ for the measured and calculated $I_{\rm p}$ evolution respectively, see table \ref{tab:comparison}. Note that in panel (b), the Dreicer generation rate is scaled by a factor 10$^{11}$ to be visible.}
    \label{fig:34183Hyb}
\end{figure}

Relevant parameters from the simulations are shown in table \ref{tab:comparison} for all four discharges, along with the measured parameters, where applicable. The CQ time listed in table \ref{tab:comparison} is defined as the time from the beginning of the $I_{\rm p}$ spike to the point where ${\rm d} I_{\rm p}/{\rm d} t < 25$ kA/ms, and the post-CQ total plasma current as the measured or calculated value of $I_{\rm p}$ at this time point. The calculated currents and RE generation rates are shown in figure \ref{fig:34183Hyb} for AUG discharge \#34183, as an example.

As explained in section \ref{sec:GOcurrent}, the phenomena resulting in the measured current spike are not modelled by \textsc{go}, and hence, the current spike is not seen in the plot of the simulated $I_{\rm p}$ evolution. In addition to this expected discrepancy, we observe that in all the simulated discharges, the simulations 
\begin{itemize}
	\item predict a partial CQ (i.e.~neither full conversion nor complete CQ),
	\item overestimate the total RE generation, leading to an underestimation of the CQ time and an overestimation of the post-CQ $I_{\rm p}$
	\item overestimate the $I_{\rm p}$ decay rate during the CQ and
	\item predict a tiny Dreicer RE seed generation, relative to the hot-tail RE seed.
\end{itemize}
These qualitative indicators of the model's performance are identical to those reported in Ref.~\cite{Insulander-JPP-20}, i.e.~we must conclude that these models need to be amended to be able to reproduce the observed $I_p$ evolution. The temperature evolution and the absence of RE losses are the main common features of these otherwise very different simulations, and in the following, we try to improve the agreement with experiment by changing to the exponential temperature model, and by modelling RE losses where appropriate.

\begin{table*}
\centering
\caption{Key parameters for the \textsc{go} simulations and the corresponding data deduced from measurements.}
\label{tab:comparison}
\begin{tabular}{l c c c c} 
\br
Discharge number							& 33108		& 34183		& 35649		& 35650 \\
\mr
Measured CQ time [ms]						& 3.8		& 4.4		& 4.7		& 4.2 \\
Calculated CQ time [ms]						& 1.1		& 1.4		& 1.3		& 1.2 \\
Measured post-CQ $I_{\rm p}$ [MA]			& 0.24		& 0.24		& 0.21		& 0.22 \\
Calculated post-CQ $I_{\rm p}$ [MA]			& 0.62		& 0.60		& 0.61		& 0.60 \\
\br
\end{tabular}
\end{table*}

\subsection{AUG simulations with the exponential temperature model}
\label{sec:Ipmatching}

As discussed in Ref.~\cite{Insulander-JPP-20}, the (almost) consequent overestimation of the post-CQ total plasma current is likely due the fact that losses of the RE seed population caused by the stochastization of the magnetic flux surfaces during the TQ are not modelled by \textsc{code} (and also not in \textsc{go}). In the \textsc{code} simulations reported in Ref. \cite{Insulander-JPP-20}, the 2D momentum distribution of the entire electron population is modelled as a continuum, i.e.~no explicit distinction is made between bulk electrons and RE. In contrast, \textsc{go} models the REs as a distinct electron population, and also models each RE generation mechanism separately. This makes it possible to model the loss of REs generated by the hot-tail mechanism by multiplying the number of hot-tail REs generated in each time step by a damping factor $f_d$. For clarity, we discuss the hot-tail seed survival fraction $f_{\rm HT} = 1-f_d$ in the following. The loss of hot-tail generated REs is expected to be significant since they are generated during the TQ when confinement is impaired due to the magnetic flux surfaces being broken up as noted in section \ref{sec:GOtemperatureExp}. The losses of REs generated through the Dreicer and avalanche mechanisms are expected to be less important since they are predominantly generated during the later phase of the disruption when the flux surfaces have re-formed.

The four AUG discharges were simulated with hot-tail losses and the hybrid temperature model, but the overprediction of the post-CQ $I_{\rm p}$ remained also when assuming loss of all hot-tail RE. When the hot-tail seed was assumed lost, the Dreicer generation increased instead, still resulting in a high total RE generation. In these cases, it was also found that the $I_{\rm p}$ evolution depended strongly on the temperature at which the switch was made from using equation \eref{eq:Texp} to invoking the energy transport equation, while the model is only predictive of the temperature evolution during the CQ if it is insensitive to $T_{\rm switch}$ in the physically motivated 10-100 eV range. Moreover, the calculated temperature in many cases increased to about 100 eV after the switch, which contradicts measurement data, as discussed in section \ref{sec:GOtemperatureHyb}. We therefore turned to the exponential temperature model, using equation \eref{eq:Texp} for the entire simulation, although other RE loss channels may also be part of the explanation for the inability to match observations.

\begin{figure}
(a)\vspace{-0.2cm}\\ \subfloat{\includegraphics[width=0.43\textwidth]{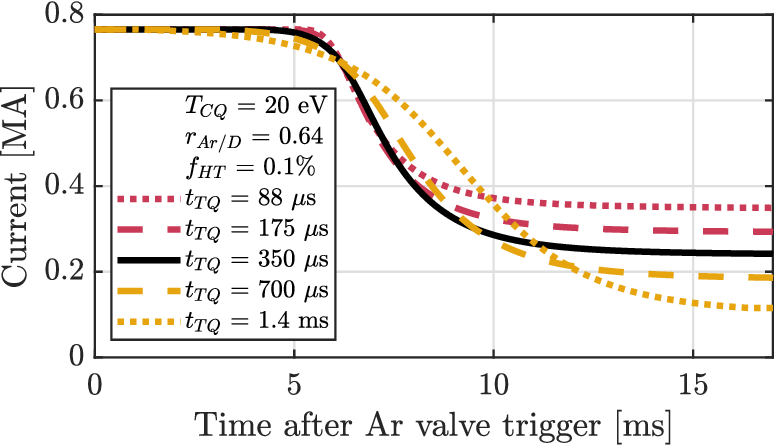}}
\vspace{-0.6cm}\\
(b)\vspace{-0.2cm}\\ \subfloat{\includegraphics[width=0.43\textwidth]{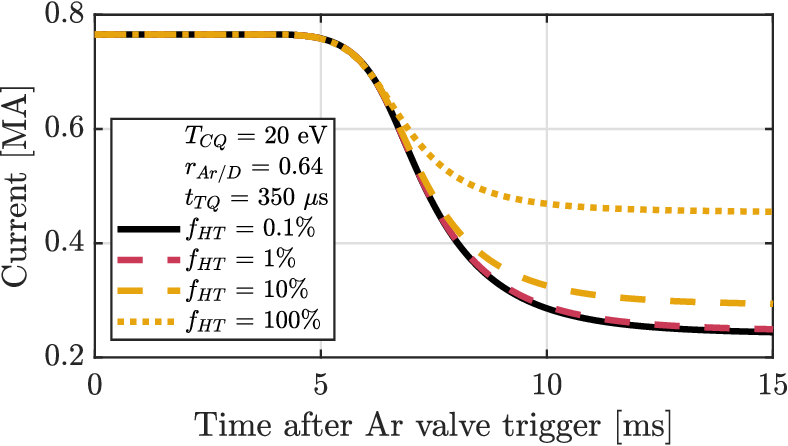}}
\vspace{-0.6cm}\\
(c)\vspace{-0.2cm}\\ \subfloat{\includegraphics[width=0.43\textwidth]{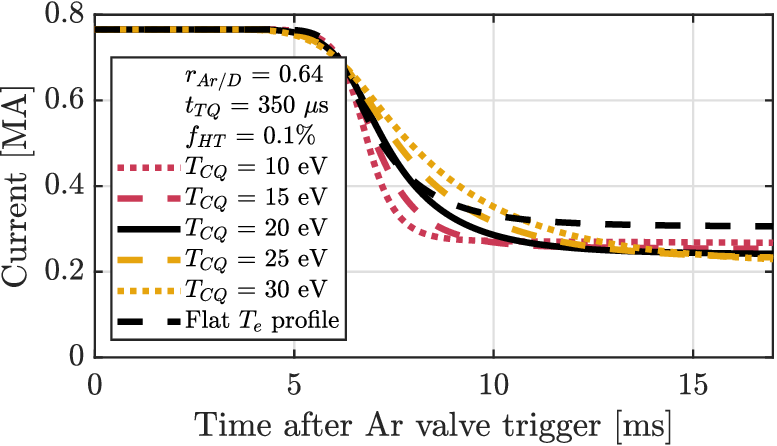}}
\vspace{-0.6cm}\\
(d)\vspace{-0.2cm}\\ \subfloat{\includegraphics[width=0.43\textwidth]{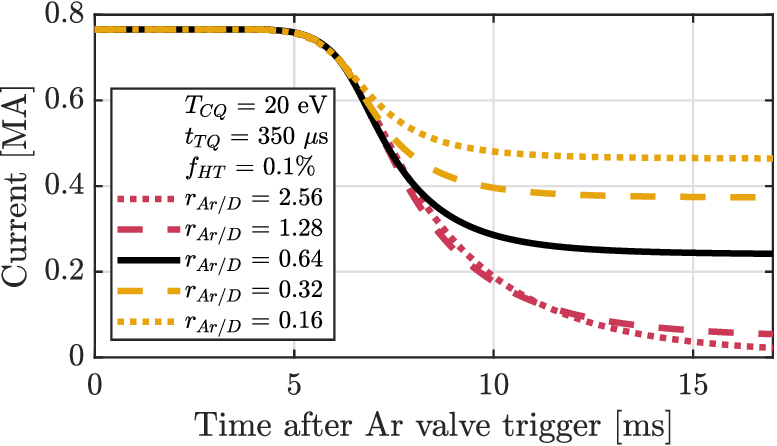}}
    \caption{Current evolution in simulations of AUG discharge \#34183, illustrating the sensitivity of the $I_p$ evolution to the simulation parameters (a) $t_{\rm TQ}$ (b) $f_{\rm HT}$, (c) $T_{\rm CQ}$ and (d) $r_{\rm Ar/D}$.  In panel (c), the sensitivity to the power profile is also illustrated by a black dashed line, showing the current evolution for $T_{\rm CQ}$ = 20 eV, but with also the initial $T_e$ profile being flat.
}
    \label{fig:scans}
\end{figure}

As mentioned in section \ref{sec:GOcurrent}, the $I_{\rm p}$ evolution is sensitive to the temperature evolution, determined by the parameters $T_{\rm initial}$, $T_{\rm CQ}$ and $t_{\rm TQ}$. $T_{\rm initial}$ can be measured and is thus taken from experimental data, but $t_{\rm TQ}$ and $T_{\rm CQ}$ are regarded as free parameters. The hot-tail RE generation is exponentially sensitive to $t_{\rm TQ}$ \cite{Smith-PoP-08}, with a quick TQ (small $t_{\rm TQ}$) resulting in a large hot-tail RE generation early in the disruption and consequently a higher post-CQ $I_{\rm p}$. An increased $t_{\rm TQ}$ results in a smaller RE generation, see figure \ref{fig:scans}(a). The hot-tail seed loss parameter can obviously be adjusted to counteract the effect of a small $t_{\rm TQ}$, but a small $t_{\rm TQ}$ results not only in increased hot-tail RE generation, but also increased Dreicer RE generation, so the effect of a small $t_{\rm TQ}$ can not be completely cancelled by assuming that the hot-tail seed is lost. As shown in figure \ref{fig:scans}(b), variations of $f_{\rm HT}$ below 1\% have a minor impact on the results for the typical size of the hot-tail seed in these simulations, because at this point, Dreicer generation becomes the dominant RE generation mechanism.

The parameter $T_{\rm CQ}$ affects the $I_{\rm p}$ decay rate during the CQ. Since the duration of the TQ ($\sim 1$~ms) is much shorter than the duration of the CQ ($\sim 5$~ms), $T_{\rm e} \approx T_{\rm CQ}$ for most of the CQ and the plasma resistivity $\sigma_{\parallel} \propto T_e^{3/2}$ determines the $I_{\rm p}$ decay rate, see figure \ref{fig:scans}(c). We can therefore infer $T_{\rm CQ}$ by choosing it to match the initial $I_{\rm p}$ decay rate. Note that $T_{\rm CQ}$ is not necessarily the prevailing temperature during the RE plateau phase, but only during most of the CQ, i.e.~the temperature may well decay further later in the disruption. The black dashed line in figure \ref{fig:scans}(c) shows the current evolution assuming a flat temperature profile with a value corresponding to a homogeneous spreading of the initial thermal energy ($T_e=1.24\;\rm keV$). Such an assumption leads to 30\% higher final runaway current, since the runaway generation becomes more efficient in a larger volume, but the overall evolution of the current remains qualitatively the same.

The amount of Ar assimilated in the plasma, quantified by the parameter $r_{\rm Ar/D}$, has a direct effect on $I_{\rm p}$, a higher $r_{\rm Ar/D}$ leading to a lower post-CQ $I_{\rm p}$, see figure \ref{fig:scans}(d), until the post-CQ $I_{\rm p}$ approaches zero.

The search of the parameter space was done iteratively, starting from the set of parameters used in the initial hybrid temperature model simulations. The values of $r_{\rm Ar/D}$ used in the initial simulations were kept the same in these new simulations. $T_{\rm CQ}$ was then chosen to match the $I_{\rm p}$ decay rate during the CQ, defined as $\Delta I_{\rm p}/\Delta t$ during half of the CQ (while $I_{\rm p}$ decreased through 25\% to 75\% of the CQ). Then the hot-tail seed loss was adjusted to match the measured post-CQ $I_{\rm p}$. For discharges \#34183 and \#35649, the post-CQ $I_{\rm p}$ remained too high even when all hot-tail RE seed was removed, so $t_{\rm TQ}$ was increased, keeping the hot-tail seed loss at 100\%, until the the measured post-CQ $I_{\rm p}$ was matched. Since $t_{\rm TQ}$ also affects the $I_{\rm p}$ decay rate to a small extent, the process had to be iterated a few times. The conditions for matching were that $\Delta I_{\rm p}/\Delta t$ should be within 10\% of the measured value, and that the post-CQ $I_{\rm p}$ should be within 10 kA from the measured value. The chosen values of the simulation parameters are listed in table \ref{tab:parameters}.

\begin{figure}
(a)\vspace{-0.2cm}\\ \subfloat{\includegraphics[width=0.43\textwidth]{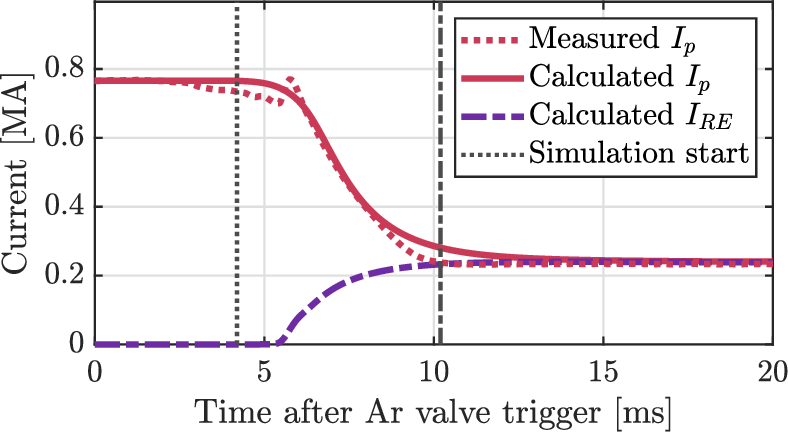}}
\vspace{-0.6cm}\\
(b)\\ \subfloat{\includegraphics[width=0.43\textwidth]{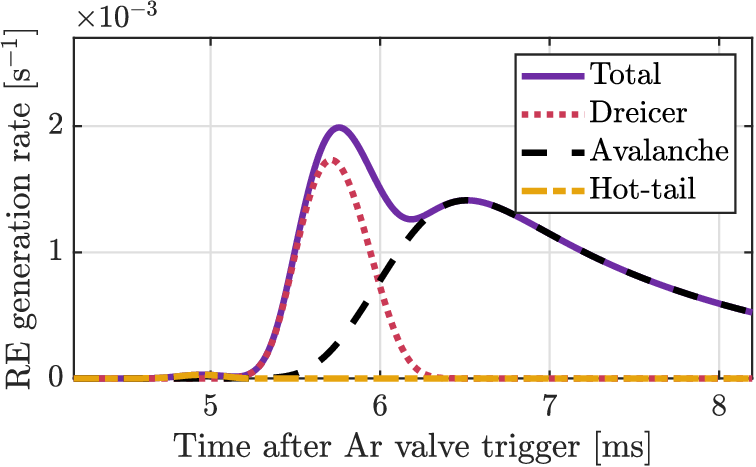}}
    \caption{(a) Current evolution and (b) RE generation rates for AUG discharge \#34183 with an exponential decay to $T_{\rm CQ}$ = 20 eV and a hot-tail RE seed loss of 99.9\%. The other parameters are $t_{TQ}$ = 0.35 ms and $r_{\rm Ar/D}$ = 0.64.}
    \label{fig:34183expdecay}
\end{figure}

Again using discharge \#34183 as an example, the $I_{\rm p}$ evolution and the RE generation rates are shown in figure \ref{fig:34183expdecay}. The corresponding figures for the other three modelled discharges are similar. The transition between the CQ and the RE plateau phase is less marked in these simulations than when the hybrid temperature model was used, since the TQ ends less abruptly when prescribed by equation \eref{eq:Texp}, as seen when comparing figure \ref{fig:34183expdecay}(a) with figure \ref{fig:34183Hyb}(a). The exponentially decaying temperature does not represent the physical $T_{\rm e}$ evolution exactly, but is a better approximation than the significant re-heating predicted when switching to the time-dependent energy transport equations as was done with the hybrid temperature model. The RE generation rates shown in figure \ref{fig:34183expdecay}(b) differ from those shown in figure \ref{fig:34183Hyb}(b) mainly in the almost completely suppressed hot-tail generation. When the hot-tail generation is suppressed, the electric field is allowed to build up, resulting in larger Dreicer generation.

\subsection{JET simulations with the hybrid temperature model}
\label{sec:AUGJETcomparison}

\begin{figure}
\includegraphics[width=0.43\textwidth]{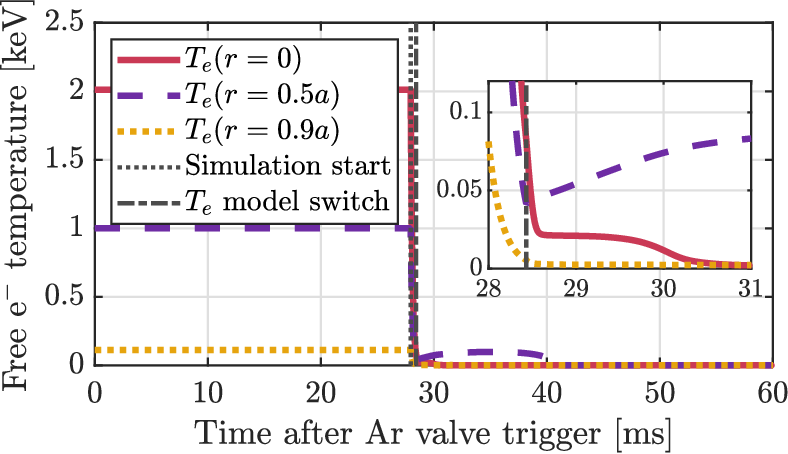}
\vspace{0.2cm}
    \caption{Temperature evolution for JET discharge \#95125 with a switch to energy transport calculations at 100 eV.}
    \label{fig:JET95125Te}
\end{figure}

\begin{figure}
(a)\vspace{-0.2cm}\\ \subfloat{\includegraphics[width=0.43\textwidth]{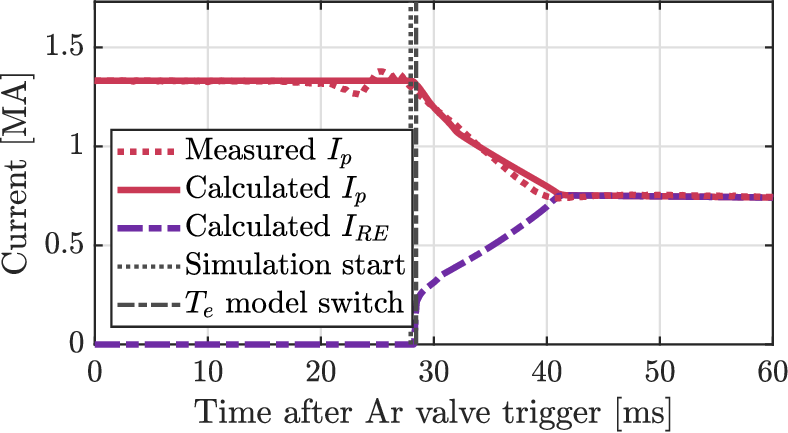}}
\vspace{-0.6cm}\\
(b)\\ \subfloat{\includegraphics[width=0.43\textwidth]{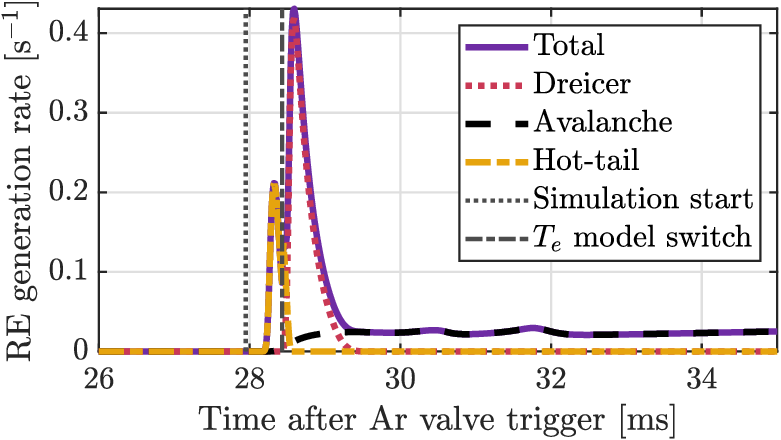}}
    \caption{(a) Current evolution and (b) RE generation rates for JET discharge \#95125 using the hybrid temperature model with a switch to energy transport calculations at 100 eV.  Note that after $\sim$ 29 ms, the RE generation is dominated by the avalanche mechanism.}
    \label{fig:JET_Ip_and_GRE_switch}
\end{figure}

Simulations of the JET disruptions were first attempted with the hybrid temperature model, but with similar complications as for the AUG discharges, i.e.~for discharges \#92454 and \#92460 no combination of parameters could be found to match the measured $I_{\rm p}$ evolution. For JET discharges \#95125 and \#95129, the energy transport calculations predicted a continued drop of the on-axis temperature from $T_{\rm switch}$ = 100 eV down to approximately 20 eV where it remained for a ms after continuing down to the final equilibrium temperature of 1 eV. This is shown for \#95125 in figure \ref{fig:JET95125Te}, where the simulation parameter values, found through an iterative search similar to that described in section \ref{sec:Ipmatching}, were $t_{\rm TQ}$ = 0.175 ms and $r_{\rm Ar/D}$ = 0.25. The resulting $I_{\rm p}$ evolution and RE generation rates are shown in figure \ref{fig:JET_Ip_and_GRE_switch}, and, as shown, the $I_{\rm p}$ evolution was reproduced with this temperature evolution. It was concluded that, again, the failure to reproduce the $I_{\rm p}$ evolution in the other discharges was mainly a consequence of a failure to predict the temperature evolution. Hence, simulations were performed with a prescribed exponential temperature drop down to a temperature of approximately 20 eV, which made it possible to reproduce the $I_{\rm p}$ evolution also for the other JET discharges listed in table \ref{tab:JETshots}.

\subsection{JET simulations with the exponential temperature model}
\label{sec:JETexponential}

As stated in section \ref{sec:JETdensities}, the fraction of the injected Ar atoms that assimilate in the plasma, $f_{\rm Ar}$, is constrained by the condition that the maximum calculated line integrated free electron density, which is very sensitive to the assimilated Ar density, should not deviate by more than 10\% from the measured value. Since the calculated free electron density is not very sensitive to the other simulation parameters, the Ar assimilation fraction, and thereby the ratio $r_{\rm Ar/D}$, was chosen first.

The other parameters were found iteratively, as for the AUG discharges. Once the assimilation fraction $f_{\rm Ar}$, and hence the values of $r_{\rm Ar/D}$ ($r_{\rm Ar/D} = f_{\rm Ar}N_{\rm Ar}/(V_{\rm p} n_{\rm e0}) = 0.5N_{\rm Ar}/(V_{\rm p} n_{\rm e0}$)),  were chosen, $T_{\rm CQ}$ was chosen to match the $I_{\rm p}$ decay rate, and the value of $t_{\rm TQ}$ was chosen to match the measured post-CQ $I_{\rm p}$. In all cases, since Dreicer was the dominant RE generation mechanism, the $I_{\rm p}$ evolution could be modelled without assuming hot-tail losses, so the hot-tail seed loss was neglected. The conditions for matching were the same as for the AUG discharges, i.e.~that $\Delta I_{\rm p}/\Delta t$ (for definition see section \ref{sec:Ipmatching}) should be within 10\% of the measured value, and that the post-CQ $I_{\rm p}$ should be within 10 kA from the measured value.

\begin{figure}
(a)\vspace{-0.2cm}\\ \subfloat{\includegraphics[width=0.47\textwidth]{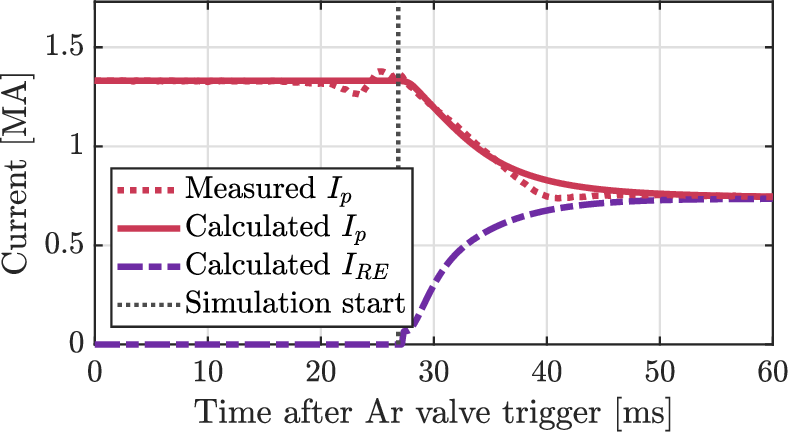}}
\vspace{-0.6cm}\\
(b)\\ \subfloat{\includegraphics[width=0.47\textwidth]{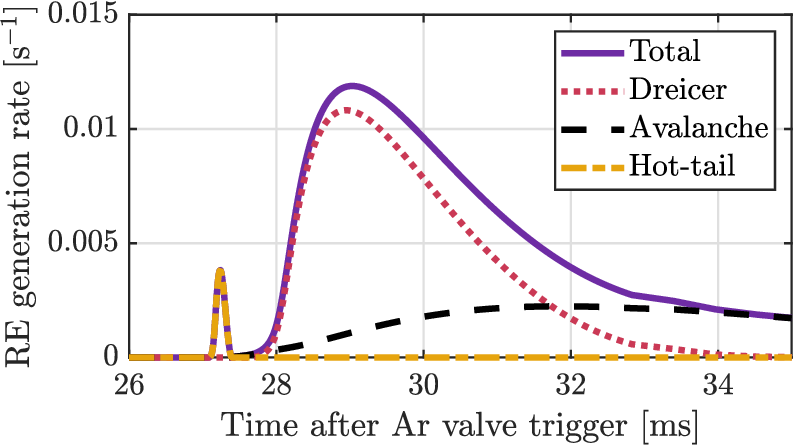}}
\vspace{-0.6cm}\\
(c)\\ \subfloat{\includegraphics[width=0.47\textwidth]{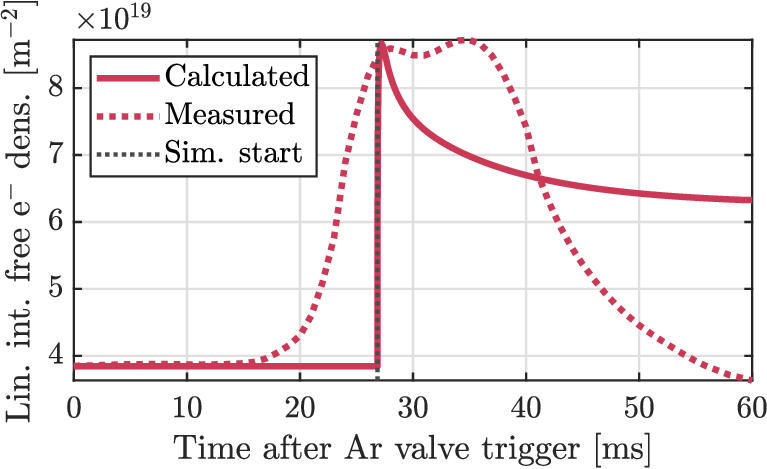}}
    \caption{(a) Current evolution, (b) RE generation rates and (c) line-integrated free electron densities for JET discharge \#95125 with exponential temperature decay to 21 eV. The reason for the decrease in the measured density after $\sim$40 ms is probably that the actual temperature has dropped to the $\sim$1 eV range in the absence of an ohmic current and associated ohmic heating, leading to lower ionization.}
    \label{fig:JET_Ip_and_GRE_expdecay}
\end{figure}

The $I_{\rm p}$ evolution shown in figure \ref{fig:JET_Ip_and_GRE_expdecay} shows a less abrupt transition from the CQ to the RE plateau phase than the measured data, just as for the AUG case in figure \ref{fig:34183expdecay}(a). This is, once again, due to the smooth end of the TQ given by the exponential approximation. The comparatively small hot-tail peak is visible to the left in figure \ref{fig:JET_Ip_and_GRE_expdecay}(b), whereas the Dreicer generation continues for several ms. This differs significantly from the generation rates shown in figure \ref{fig:JET_Ip_and_GRE_switch}(b), where the Dreicer peak is almost as narrow as the hot-tail peak.

The calculated line-integrated free electron density is shown together with a measurement-based estimate of the same quantity in figure \ref{fig:JET_Ip_and_GRE_expdecay}(c). The estimate is based on the line integrated density measured by interferometry (KG1 diagnostic, corrected for fringe jumps). The much faster increase of the calculated density immediately at the start of the simulation is due to the assumption that all Ar enters the plasma instantaneously, which is obviously not correct, but has a minor impact on the calculated plasma current.

\begin{figure}
(a)\vspace{-0.2cm}\\ \subfloat{\includegraphics[width=0.47\textwidth]{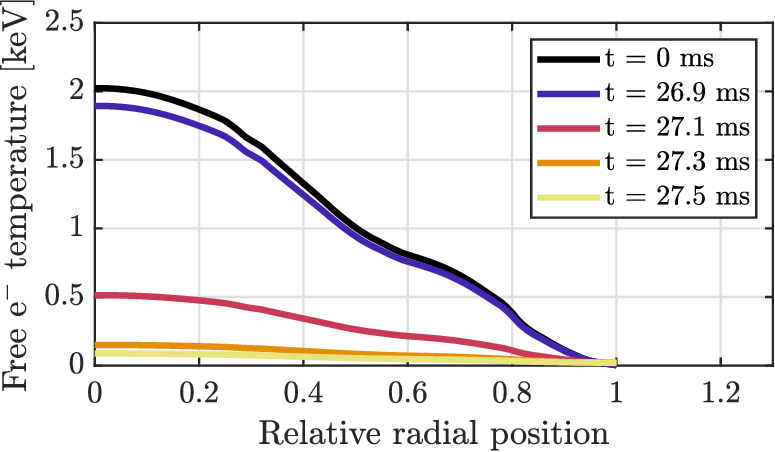}}
\vspace{-0.6cm}\\
(b)\\ \subfloat{\includegraphics[width=0.47\textwidth]{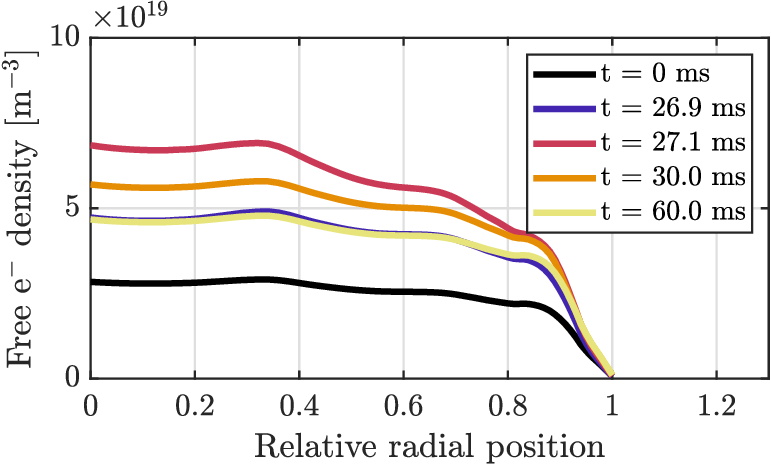}}
    \caption{Radial profiles of the free electron (a) temperature and (b) density.}
    \label{fig:profiles}
\end{figure}

Using only the exponential temperature model, the free electron temperature profile remains the same throughout the simulation, as shown in figure \ref{fig:profiles}(a). The free electron density profile is re-calculated to be consistent with the temperature profile, as the ionization states of the injected argon changes with the temperature. Since the radial distribution of the injected argon is approximated by the initial distribution of the deuterium, the density profile remains nearly constant throughout the simulation, as shown in figure \ref{fig:profiles}(b). Some small changes in the profile shape can be seen when comparing the density profile at 26.9 ms and at the end of the simulation (60.0 ms).

\begin{table*}
\centering
\caption{Parameters used in the simulations run with the exponential temperature model. For the JET cases, the hot-tail survival fraction is unconstrained, since the $I_p$ evolution could be reproduced without assuming hot-tail losses.}
\label{tab:parameters}
\begin{tabular}{l c c c c} 
\br
 									& Argon-to-deuterium	& Thermal quench& Hot-tail 		& Post-TQ		\\	
 									& density ratio 		& time parameter & survival fraction& temperature	\\				
									& $r_{\rm Ar/D}$	& $t_{\rm TQ}$	& $f_{\rm HT}$	& $T_{\rm CQ}$  \\
									& 					& ms 			& \% 			& eV 			\\
\mr
\multicolumn{5}{l}{\textit{AUG discharges}}																\\
\mr
33108								& 1.40				& 0.152 			& 0.8			& 17			\\
34183								& 0.64				& 0.35			& $\le$ 0.1		& 20			\\
35649								& 0.90				& 0.35			& $\le$ 0.1		& 20			\\
35650								& 0.98				& 0.177			& 0.1			& 23			\\
\mr
\multicolumn{5}{l}{\textit{JET discharges}}																	\\
\mr
92454								& 0.23				& 0.150 			& -				& 19			\\
92460								& 0.16				& 0.090 			& - 				& 24			\\
95125 								& 0.11				& 0.180 			& - 				& 21 			\\
95129 								& 0.12				& 0.085 			& - 				& 24 			\\
\br
\end{tabular}
\end{table*}

%%%%%%%%%%% Discussion and conclusions
\section{Discussion and conclusions} \label{sec:conclusions}

The most important parameters for the simulated discharges in JET and AUG are listed in table \ref{tab:tokamaks} and the simulation parameters in table \ref{tab:parameters}. We note that the initial free electron densities and also the externally applied magnetic field are similar for the two sets of discharges, whereas the machine size and the initial plasma current are significantly larger in JET. It should be noted that the higher current in JET is partly a consequence of the larger machine size - the current densities are similar (1.5 -- 1.8 MA/m$^2$ for AUG and 1.9 -- 2.0 MA/m$^2$ for JET). In AUG, the initial temperature and the injected Ar density ($r_{\rm Ar/D} \cdot n_{e0}$) are larger than in the JET discharges. 

The simulation parameters needed to reproduce the $I_{\rm p}$ evolution when using the same modelling strategy are similar, except that the hot-tail seed survival fraction had to be chosen very small to reproduce the AUG discharges, whereas its value was unimportant for modelling of the JET discharges, due to the rather small hot-tail generation. The argon assimilation rates (expressed through the argon-to-deuterium density ratio $r_{\rm Ar/D}$) are based on experimental data as discussed in section \ref{sec:GOdensities}, and are consistently larger in the AUG discharges.

Our simulations indicate that the hot-tail RE generation is much smaller in JET as compared with AUG. This may be understood by the fact that at a fixed TQ time, the hot-tail seed increases exponentially with initial temperature due to the longer slowing-down time of the hot-tail electrons. According to the simulations, the TQ times in the two devices appear to be similar, although this property would be expected to scale with machine size \cite{Hender_2007}. This may be because the higher initial temperature in the AUG discharges is compensated by a relatively higher density of impurities (due to the smaller plasma volume), which can then radiate the heat more efficiently. 

An important remaining problem to be solved before reliable predictive simulations can be made is the self-consistent modelling of the temperature evolution. Although we found that a time-dependent energy transport model, including ohmic heating and impurity radiation losses, allowed for reproduction of the plasma current evolution for some JET discharges, this model was prone to predict a re-heating of the plasma after a forced temperature drop in other discharges in a way that contradicts the experimental data. This suggests that some additional losses are present, such as radiation from wall impurities and transport losses remaining during the CQ. The importance of transport losses is expected to be higher in a smaller machine, which might explain why the prediction of an experimentally excluded re-heating was less common for JET than for AUG. Time dependent radial profiles of the magnetic field fluctuations are difficult to obtain from experiment, although their amplitude during the current quench has often been found to be fairly small \cite{Gill}. Numerical modelling of the transport due to magnetic fluctuations \cite{Svensson_2020} would therefore require exploration of a fairly large additional parameter space and the effect of this on the results presented remains an open question.

The parameter $T_{\rm CQ}$ used in the exponential temperature model lies consistently around 20 eV for both machines. It is assumed that whereas this temperature prevails during most of the CQ, it is only maintained during this period through collisional heating by the remaining ohmic current, and the temperature during the RE plateau is likely to be about 1 eV on average. The existence of such a temperature plateau is predicted by the energy transport equations in the cases when the $I_{\rm p}$ evolution is well reproduced by this model. Experimental data can neither confirm nor exclude the existence of such a temperature plateau.

The main conclusion of the presented set of simulations is that the RE generation models in \textsc{go} are able to reproduce experimentally measured runaway currents for both AUG and JET using similar assumptions and modelling strategies. This increases our confidence that the RE models adequately describe the runaway physics in tokamaks of different sizes. The relative simplicity of the code makes it possible to run a large number of simulations and qualitatively investigate the consequences of the simultaneous variation of several different parameters over large intervals.

%An important conclusion of the presented set of simulations is that the fluid models implemented in \textsc{go}, and the implemented RE generation models, are able to reproduce experimentally measured total plasma currents for both AUG and JET using similar assumptions and modelling strategies. This increases our confidence that the models adequately describe the disruption physics, and suggest that \textsc{go} can be used for simulations of tokamaks of different sizes. The relative simplicity of the code makes it possible to run a large number of simulations and qualitatively investigate the consequences of the simultaneous variation of several different parameters over large intervals.

    \section*{Acknowledgements}
The authors are grateful to L Hesslow, M Hoppe, I Svenningsson, I Pusztai, A Boboc and G Pautasso for fruitful discussions. This work has been carried out within the framework of the EUROfusion Consortium and has received funding from the Euratom research and training programme 2014 - 2018 and 2019 - 2020 under grant agreement No 633053 and from the European Research Council (ERC) under the European Union's Horizon 2020 research and innovation programme under grant agreement No 647121. The views and opinions expressed herein do not necessarily reflect those of the European Commission.  The work was also supported by the Chalmers Gender Initiative for Excellence (Genie), the Swedish Research Council (Dnr.~2018-03911) and the EUROfusion - Theory and Advanced Simulation Coordination (E-TASC). 

\newcommand{\newblock}{}
\bibliographystyle{unsrt}
\bibliography{plasmabibliography}

\end{document}